\documentclass[useAMS,usenatbib]{mn2e}

\usepackage{times}
\usepackage{graphicx}
\usepackage{rotating}

\title[Proper motions of field L and T dwarfs]{Proper motions of field L and T dwarfs}
\author[R. F. Jameson et al.]{R. F. Jameson$^{1}$\thanks{E-mail:
rfj@star.le.ac.uk}, S. L.Casewell$^{1}$, N. P. Bannister$^{1}$ , N. Lodieu $^{2,1}$, K. Keresztes$^{1}$,
\newauthor P. D. Dobbie $^{3,1}$, S. T. Hodgkin $^{4}$\\
$^{1}$Department of Physics and Astronomy, Univeristy of Leicester, Univeristy Road, Leicester, LE1 7RH, UK\\
$^{2}$Instituto de Astrof\'isica de Canarias, V\'ia L\'actea s/n, E-38205 La Laguna, Tenerife, Spain\\
$^{3}$Anglo-Australian Observatory, PO Box 296, Epping NSW 1710 Australia\\
$^{4}$CASU, Institute of Astronomy,University of Cambridge, Maddingley Road, Cambridge, CB3 0HA, UK \\}
\begin{document}

\date{}

\pagerange{\pageref{firstpage}--\pageref{lastpage}} \pubyear{2007}

\maketitle

\label{firstpage}

\begin{abstract}
The proper motion measurements for 143 previously known L and T dwarfs are presented. From this sample we identify and 
discuss 8 high velocity L dwarfs.  We also find 4 new wide common proper motion binaries/multiple systems.
Using the moving cluster methods we have also identified 
a number of L dwarfs that may be members of the Ursa Major (age $\approx$ 400 Myr), the Hyades (age $\approx$ 625 Myr) and the 
Pleiades (age $\approx$ 125 Myr) moving groups.
\end{abstract}


\section{Introduction}
Brown dwarfs may be thought of as failed stars. These low mass ($\leq$70 M$_{\rm Jup}$ \citet{burrows01}), cool objects are the lowest mass objects that 
the star formation process can produce. The majority of the brown dwarfs that have been discovered to date are field objects discovered using surveys such as the Two Micron All Sky Survey (2MASS; \citet{skrutskie06}, see \citet{leggett02} for examples), the DEep Near-Infrared Sky survey (DENIS; \citet{denis05}, see \citet{delfosse99} for examples), the Sloan Digital Sky Survey (SDSS;\citet{york00} see \citet{hawley02} for examples) and the UKIRT Deep Infrared Sky Survey (UKIDSS; \citet{lawrence06}, see \citet{kendall07} for examples). 
However, to study brown dwarfs in depth, a knowledge of their age is essential, which means we must study brown dwarfs in open star clusters or moving groups.

Once a brown dwarf has been proved to belong to an open star cluster, or a moving group, then the age of the dwarf is known, allowing meaningful comparisons to evolutionary models to be made. The most recent example of this is the study done by \citet{bannister07} who used existing proper motions and parallax measurements to show that a selection of field dwarfs in fact belong to the Ursa Major and Hyades moving groups. The importance of this study, is that these are the first brown dwarfs to be associated with an older cluster or group. Older clusters such as the Hyades are expected to contain very few or no brown dwarfs or low mass members, due to the dynamical evolution of the cluster over time \citep{adams02}. However, these escaped low mass objects may remain members of the much larger moving group that surrounds the cluster. 

To continue the  study started by \citet{bannister07}, proper motions need to be measured for the majority of the field brown dwarfs currently known.
This has been accomplished using the wide field camera (WFCAM, \citet{casali07}) of the United Kingdom Infrared Telescope (UKIRT). Using these WFCAM images and existing data  we have measured proper motions for 143 L and T dwarfs listed in the online Dwarf Archive\footnote{See
http://spider.ipac.caltech.edu/staff/davy/ARCHIVE/, a webpage
dedicated to L and T dwarfs maintained by C.\ Gelino, D.\ Kirkpatrick, 
and A.\ Burgasser.}. 

This proper motion data may be put to a number of uses. Taken with measured radial velocities and distances, it can yield all three components 
of velocity (U,V,W). Using reduced proper motion diagrams it can be used as an approximate measure of distance.
However we have no radial velocities for these objects. These proper motion measurements can however also 
be used to help identify objects as members of a star cluster or members of a moving group via the moving cluster method.

Our proper motion data is discussed and listed in section 2 of this paper.
From the proper motion measurements, we find 5 new wide common proper motion binaries/multiple systems. We also identify 8 high velocity L dwarfs, which are discussed in section 4. We suggest that these L dwarfs are probably old and belong to the thick disc or halo population of the galaxy. This in turn suggests that they are likely to be very faint stars rather than brown dwarfs.
Finally in section 5 we identify a number of L and T dwarfs that may be members of the Hyades, Pleiades and Ursa Major moving groups .

\section{Proper Motion Measurements}
\subsection{Data acquisition and reduction}
In order to measure proper motions for known L and T dwarfs we observed  143 L and T dwarfs from the DwarfArchive (detailed in table \ref{pm}, see http://spider.ipac.caltech.edu/staff/davy/ARCHIVE/ for discovery references) with declinations between -30$^{\circ}$ and 60$^{\circ}$ and \textit{J} band magnitudes of less than 16.5. 
The images were taken using WFCAM on UKIRT over the period of February 2006 to August 2006.
WFCAM is a near infrared imager consisting of 4 Rockwell Hawaii-II (HgCdTe 2048x2048) arrays arranged
 such that 4 separately pointed observations can be tiled together to cover a filled square of sky covering 0.75 square degrees 
with 0.4 arcsecond pixels \citet{casali07}. However, as we only required the image of the brown dwarf in question, we only used array 3  
which is regarded as the least noisy array. WFCAM is ideal for this work, as the large field of view per chip means there are many other 
stars in the image, which can be used as astrometric reference stars.
The images were taken in the \textit{J} band in non-photometric conditions using exposure times of $\approx$5-10 minutes and a nine point dither pattern. These exposure times gave S/N$\approx$100 even in the poor conditions.

The images were reduced at the Cambridge Astronomical Survey Unit (CASU) using procedures which have been custom written 
for the treatment of WFCAM data (Irwin et al., in preparation, \citet{dye06}). In brief, each frame was 
debiased, dark corrected and then flat fielded. The individual dithered images were stacked before having an object detection 
routine run on them. 
The frames were astrometrically calibrated using point sources in the
Two micron All Sky Survey (2MASS) catalogue. The accuracy is typically $\approx$0.1'' \citep{dye06} 
The photometric calibration
employed by the CASU pipeline also relies on 2MASS data (there are typically hundreds of 2MASS calibrators per detector) and is found to be
accurate to $\approx$2\% in good conditions \citep{warren07},  however as we wished to measure proper motions, the astrometric calibration was more important than the photometric calibration for these data.

\subsection{Calculating proper motions}
The astrometry for 2MASS is good to 80 mas over the whole survey, and to 50 mas over a small area \citep{skrutskie06}. 
Because the WFCAM astrometry is also calibrated to the 2MASS catalogue, accurate relative proper motion measurements could be calculated 
simply by taking the difference in 2MASS and WFCAM positions and dividing by the epoch difference. 
We calculated the epoch difference  by taking the difference in the Julian date as given in the FITS header for each image, which is between 5 and 9  years with the average epoch difference being 7.1 years.
\citet{lodieu07} employ a similar method for calculating proper motions using the UKIDSS Galactic Cluster Survey, also using WFCAM and 2MASS. 

The proper motion measurements for each object in every WFCAM image (i.e. array 3) were calculated by this method. 
This motion, was then converted into mas yr$^{-1}$. The proper motion has been calculated directly from the RA and dec of the object in question, not 
from pixel motion on images, hence $\mu_{\alpha}$=($\Delta\alpha$/$\Delta$T)$\cos\delta$ `` yr$^{-1}$ if $\Delta\alpha$ is converted to arcseconds. 
These proper motions are relative proper motions, in the sense that they are relative to the bulk of the background stars in the field, which are generally moving slowly enough to be assumed to have zero motion. 
However, we checked the reference star motion  so that the proper motion of the brown dwarf could be altered if there was a standard offset in the field.

\begin{figure*}
\begin{center}
\scalebox{0.60}{\includegraphics[angle=270, width=\linewidth]{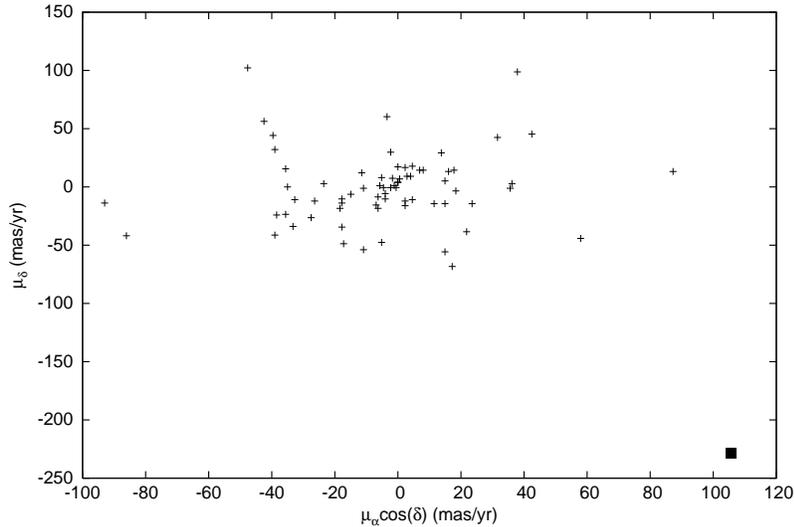}}
\caption{
\label{pmplot1}
Proper motion vector diagram showing the case where there is a large spread in the reference stars compared to the dwarf motion
i.e. they are not all concentrated around zero. The brown dwarf J1338+04 (L1, Reid et al., in prep., $\mu_{\alpha}$cos$\delta$=111.65$\pm$13.88, $\mu_{\delta}$=-224.27$\pm$12.50) is marked by the filled square.}
\end{center}
\end{figure*}

\begin{figure*}
\begin{center}
\scalebox{0.60}{\includegraphics[angle=270, width=\linewidth]{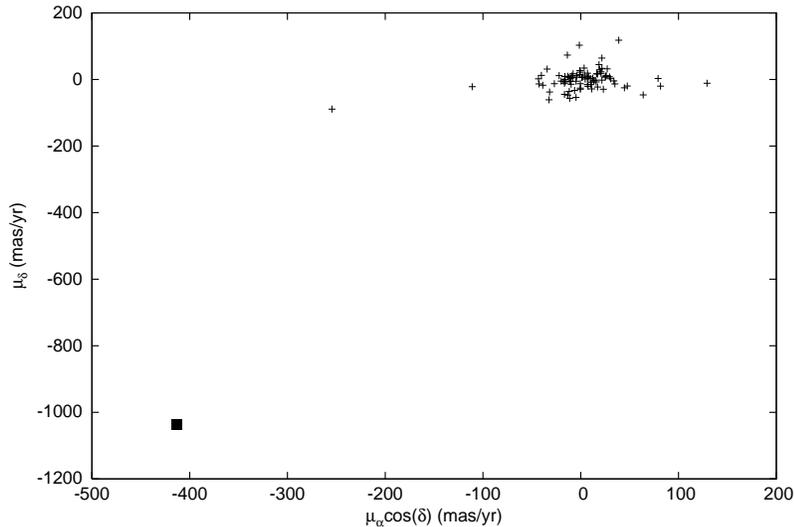}}
\caption{
\label{pmplot3}
Proper motion vector diagram showing the ideal case. The reference stars are concentrated on zero compared to the dwarf motion, and the brown dwarf is marked by the filled square(J1331-01, L6, \citet{hawley02}, $\mu_{\alpha}$cos$\delta$=-407.10$\pm$18.61, $\mu_{\delta}$=-1029.84$\pm$14.25) has a large proper motion.}
\end{center}
\end{figure*}

The proper motions were separated in $\mu_{\alpha}cos\delta$ and $\mu_{\delta}$ from -500 to 500 mas yr$^{-1}$ in each direction, in bins of size 20 mas yr$^{-1}$, and the number of objects 
falling into each bin were totalled. 
We then fitted a two dimensional Gaussian to the data for each field to determine the spread of the reference stars, as well as the true centre of the motion. The process was then repeated after the initial fit, rejecting any objects that lay outside 3$\sigma$ of the fitted Gaussian,
 before fitting another Gaussian to this data. 
This fitting was important in some cases as the reference stars had quite a large spread, and in other cases the proper motion of our brown dwarf
was of the same order of magnitude as the references. This is illustrated by figures \ref{pmplot1} and \ref{pmplot3}.
These centroiding error, or the centres of the fit were then subtracted from the calculated proper motion measurements.

\citet{lodieu07} assumed the errors on their proper motions to be $\approx$ 10 mas yr$^{-1}$, but no formal error calculation was made.
We used the $\sigma$ value of the Gaussian to determine the error on our measurements. In general the errors were of the order of $\approx$ 15 mas yr$^{-1}$. The quoted errors are the $\sigma$ value of 
the Gaussian fitted to the proper motion points for each image. Strictly this should be the $\sigma$ plus the position error of the object added in quadrature.
However, 
the centroiding errors for WFCAM are less than 2 mas yr$^{-1}$, and so are small compared to the $\sigma$ value.  
Table \ref{pm} shows the 2MASS name, RA, dec, proper motion and associated errors for the 143 objects studied. Their discovery papers may be found from the DwarfArchive. All of the objects in table \ref{pm} are 2MASS objects and  are identified by their 2MASS name. Hereafter they are referred to as J followed by the first 4 digits, the plus or minus sign, and the next two digits.

 \begin{table*}
\caption{\label{pm} 2MASS Name, RA, Dec,$\mu_{\alpha}$cos$\delta$ and $\mu_{\delta}$ for all of the L and T dwarfs for which we measured proper motions. }
\begin{center}
\begin{tabular}{l r r r r}
\hline
Name&RA&Dec&$\mu_{\alpha}$cos$\delta$&$\mu_{\delta}$\\
2MASS&\multicolumn {2}{|c|}{J2000}&\multicolumn {2}{|c|}{mas yr$^{-1}$}\\
\hline
J00001354+2554180 & 00 00 13.54 & 25 54 18.94 & 5.84 $\pm$ 19.47 & 130.03 $\pm$ 21.94\\
J00011217+1535355 & 00 01 12.23 & 15 35 34.45 & 149.65 $\pm$ 22.72 & $-$169.14 $\pm$ 14.85\\
J00100009-2031122 & 00 10 00.15 & $-$20 31 12.10 & 117.47 $\pm$ 19.59 & 30.74 $\pm$ 17.06\\
J0015447+351603 & 00 15 44.80 & 35 16 00.96 & 83.19 $\pm$ 17.57 & $-$245.65 $\pm$ 13.80\\
J00191165+0030176 & 00 19 11.64 & 00 30 17.46 & $-$28.79 $\pm$ 18.25 & $-$25.01 $\pm$ 11.26\\
J0030438+313932 & 00 30 43.81 & 31 39 31.56 & $-$49.44 $\pm$ 15.89 & $-$56.86 $\pm$ 13.50\\
J00320509+0219017 & 00 32 05.26 & 02 18 59.82 & 417.19 $\pm$ 16.58 & $-$311.77 $\pm$ 12.66\\
J0032431-223727 & 00 32 43.14 & $-$22 37 27.14 & 115.06 $\pm$ 26.78 & 20.23 $\pm$ 14.88\\
J00384398+1343395 & 00 38 44.01 & 13 43 39.11 & 63.65 $\pm$ 16.16 & $-$40.97 $\pm$ 8.33\\
J00412179+3547133 & 00 41 21.74 & 35 47 13.18 & $-$101.90 $\pm$ 16.59 & $-$16.66 $\pm$ 13.01\\
J00415453+1341351 & 00 41 54.46 & 13 41 34.35 & $-$174.28 $\pm$ 23.61 & $-$137.56 $\pm$ 36.03\\
J00452143+1634446 & 00 45 21.58 & 16 34 44.46 & 384.66 $\pm$ 17.12 & $-$26.35 $\pm$ 12.27\\
J00464841+0715177 & 00 46 48.45 & 07 15 17.44 & 98.06 $\pm$ 21.53 & $-$50.63 $\pm$ 10.04\\
J01033203+1935361 & 01 03 32.22 & 19 35 36.37 & 304.63 $\pm$ 16.54 & 35.18 $\pm$ 14.32\\
J02281101+2537380 & 02 28 11.15 & 25 37 37.84 & 256.53 $\pm$ 14.33 & $-$13.98 $\pm$ 16.88\\
J02511490-0352459 & 02 51 15.48 & $-$03 53 00.02 & 1128.19 $\pm$ 13.32 & $-$1826.44 $\pm$ 19.76\\
J09083803+5032088 & 09 08 37.73 & 50 32 05.26 & $-$394.92 $\pm$ 26.21 & $-$477.67 $\pm$ 14.92\\
J09440279+3131328 & 09 44 02.81 & 31 31 32.46 & 42.96 $\pm$ 15.82 & $-$32.53 $\pm$ 13.47\\
J10042066+5022596 & 10 04 20.55 & 50 22 58.18 & $-$133.22 $\pm$ 39.63 & $-$185.08 $\pm$ 15.42\\
J10101480-0406499 & 10 10 14.64 & $-$04 06 49.81 & $-$321.11 $\pm$ 15.90 & 20.12 $\pm$ 13.28\\
J10170754+1308398 & 10 17 07.57 & 13 08 39.05 & 60.54 $\pm$ 21.83 & $-$93.74 $\pm$ 12.30\\
J10220489+0200477 & 10 22 04.81 & 02 00 45.11 & $-$173.19 $\pm$ 18.24 & $-$398.05 $\pm$ 15.76\\
J10292165+1626526 & 10 29 21.86 & 16 26 49.60 & 358.65 $\pm$ 14.91 & $-$348.21 $\pm$ 17.05\\
J10352455+2507450 & 10 35 24.44 & 25 07 42.68 & $-$181.48 $\pm$ 16.86 & $-$271.82 $\pm$ 23.18\\
J10440942+0429376 & 10 44 09.41 & 04 29 38.20 & $-$1.87 $\pm$ 12.11 & 94.47 $\pm$ 10.49\\
J10452400-0149576 & 10 45 23.76 & $-$01 49 57.64 & $-$494.98 $\pm$ 17.78 & 11.75 $\pm$ 12.05\\
J10473109-1815574 & 10 47 30.89 & $-$18 15 57.10 & $-$347.24 $\pm$ 17.09 & 54.03 $\pm$ 14.11\\
J10484281+0111580 & 10 48 42.62 & 01 11 56.71 & $-$442.18 $\pm$ 13.09 & $-$208.76 $\pm$ 12.19\\
J10511900+5613086 & 10 51 18.82 & 56 13 06.65 & $-$231.49 $\pm$ 34.15 & $-$288.05 $\pm$ 14.40\\
J10595138-2113082 & 10 59 51.45 & $-$21 13 09.52 & 134.32 $\pm$ 17.06 & $-$158.27 $\pm$ 13.58\\
J11040127+1959217 & 11 04 01.30 & 19 59 22.56 & 74.75 $\pm$ 14.65 & 138.77 $\pm$ 20.26\\
J11101001+0116130 & 11 10 09.91 & 01 16 11.50 & $-$243.29 $\pm$ 20.82 & $-$237.94 $\pm$ 17.78\\
J11131694-0002467 & 11 13 16.96 & 00 02 46.82 & 42.12 $\pm$ 21.73 & 5.91 $\pm$ 13.39\\
J11235564+4122286 & 11 23 55.56 & 41 22 28.09 & $-$110.47 $\pm$ 25.23 & $-$60.19 $\pm$ 13.22\\
J11455714+2317297 & 11 45 57.23 & 23 17 29.31 & 154.68 $\pm$ 16.47 & $-$55.55 $\pm$ $-$6.46\\
J11480423+0254057 & 11 48 04.24 & 02 54 05.41 & 25.88 $\pm$ 24.78 & $-$41.17 $\pm$ 10.28\\
J11480502+0203509 & 11 48 05.12 & 02 03 48.88 & 237.33 $\pm$ 25.57 & $-$322.40 $\pm$ 13.25\\
J11533966+5032092 & 11 53 39.72 & 50 32 09.60 & 83.93 $\pm$ 22.95 & 60.19 $\pm$ 10.88\\
J11550087+2307058 & 11 55 00.89 & 23 07 05.96 & 25.80 $\pm$ 24.88 & 37.70 $\pm$ 20.68\\
J11593850+0057268 & 11 59 38.51 & 00 57 26.85 & 12.36 $\pm$ 22.82 & 6.65 $\pm$ 17.26\\
J12035812+0015500 & 12 03 57.61 & 00 15 48.33 & $-$1209.29 $\pm$ 18.27 & $-$260.69 $\pm$ 14.75\\
J12043036+3212595 & 12 04 30.40 & 32 12 59.37 & 91.42 $\pm$ 37.62 & $-$17.54 $\pm$ 19.03\\
J12074717+0244249 & 12 07 46.96 & 02 44 25.72 & $-$498.04 $\pm$ 17.54 & 137.51 $\pm$ 19.27\\
J12130336-0432437 & 12 13 03.18 & $-$04 32 43.97 & $-$353.58 $\pm$ 16.21 & $-$12.40 $\pm$ 12.42\\
J12212770+0257198 & 12 21 27.65 & 02 57 19.71 & $-$114.73 $\pm$ 30.13 & $-$17.60 $\pm$ 27.17\\
J12321827-0951502 & 12 32 18.20 & $-$09 51 50.97 & $-$156.29 $\pm$ 13.04 & $-$100.94 $\pm$ 16.93\\
J12392727+5515371 & 12 39 27.39 & 55 15 37.31 & 160.87 $\pm$ 29.34 & 38.06 $\pm$ 5.62\\
J12464678+4027150 & 12 46 46.85 & 40 27 14.45 & 145.05 $\pm$ 11.66 & $-$78.75 $\pm$ 16.59\\
J12565688+0146163 & 12 56 56.80 & 01 46 16.14 & $-$182.82 $\pm$ 13.21 & $-$17.05 $\pm$ 17.81\\
J12573726-0113360 & 12 57 37.30 & $-$01 13 36.18 & 81.10 $\pm$ 20.22 & $-$2.79 $\pm$ 11.02\\
J13015465-1510223 & 13 01 54.61 & $-$15 10 22.94 & $-$69.22 $\pm$ 11.67 & $-$74.28 $\pm$ 16.06\\
J13054106+2046394 & 13 05 41.05 & 20 46 39.95 & $-$23.27 $\pm$ 17.49 & 73.37 $\pm$ 26.87\\
J13120707+3937445 & 13 12 07.00 & 39 37 44.61 & $-$96.57 $\pm$ 23.23 & 13.43 $\pm$ 18.33\\
J13153094-2649513 & 13 15 30.53 & $-$26 49 53.65 & $-$677.53 $\pm$ 15.92 & $-$280.48 $\pm$ 14.66\\
J13204427+0409045 & 13 20 44.06 & 04 09 05.90 & $-$483.36 $\pm$ 19.44 & 210.86 $\pm$ 17.28\\
J13233597-1806379 & 13 23 35.92 & $-$18 06 38.14 & $-$86.89 $\pm$ 15.41 & $-$22.18 $\pm$ 18.70\\
J13262009-2729370 & 13 26 19.90 & $-$27 29 37.30 & $-$363.71 $\pm$ 16.48 & $-$15.80 $\pm$ 13.80\\
J13313310+3407583 & 13 31 32.92 & 34 07 57.20 & $-$352.59 $\pm$ 18.85 & $-$169.29 $\pm$ 18.43\\
J13314894-0116500 & 13 31 48.74 & $-$01 16 57.65 & $-$407.10 $\pm$ 18.61 & $-$1029.84 $\pm$ 14.25\\
J13322863+2635079 & 13 32 28.53 & 26 35 08.22 & $-$151.50 $\pm$ 12.41 & 40.80 $\pm$ 16.67\\
J13340623+1940351 & 13 34 06.19 & 19 40 36.01 & $-$57.97 $\pm$ 12.32 & 98.09 $\pm$ 16.04\\
J13364062+3743230 & 13 36 40.49 & 37 43 22.52 & $-$200.46 $\pm$ 8.54 & $-$60.43 $\pm$ 13.81\\
\hline
\end{tabular}
\end{center}
\end{table*}
\addtocounter{table}{-1}
 \begin{table*}
\caption{continued}
\begin{center}
\begin{tabular}{l r r r r}
\hline
Name&RA&Dec&$\mu_{\alpha}$cos$\delta$&$\mu_{\delta}$\\
2MASS&\multicolumn {2}{|c|}{J2000}&\multicolumn {2}{|c|}{mas yr$^{-1}$}\\
\hline
J13382615+4140342 & 13 38 26.04 & 41 40 31.66 & $-$153.13 $\pm$ 24.17 & $-$310.98 $\pm$ 24.70\\
J13384944+0437315 & 13 38 49.49 & 04 37 30.13 & 111.65 $\pm$ 13.88 & $-$224.27 $\pm$ 12.50\\
J13422362+1751558 & 13 42 23.57 & 17 51 55.59 & $-$70.09 $\pm$ 11.13 & $-$1.92 $\pm$ 8.31\\
J13431670+3945087 & 13 43 16.46 & 39 45 09.59 & $-$343.16 $\pm$ 31.26 & 115.99 $\pm$ 24.80\\
J13484591+0353545 & 13 48 46.00 & 03 53 52.28 & 206.90 $\pm$ 22.82 & $-$355.69 $\pm$ 11.19\\
J14023175+0148301 & 14 02 31.66 & 01 48 30.19 & $-$232.37 $\pm$ 13.59 & 7.60 $\pm$ 11.06\\
J14044167+0235501 & 14 04 41.70 & 02 35 48.51 & 53.77 $\pm$ 17.37 & $-$248.28 $\pm$ 13.22\\
J14044495+4634297 & 14 04 44.79 & 46 34 30.74 & $-$231.37 $\pm$ 30.27 & 142.97 $\pm$ 26.38\\
J14075361+1241099 & 14 07 53.47 & 12 41 10.41 & $-$312.48 $\pm$ 19.16 & 81.69 $\pm$ 18.75\\
J14111735+3936363 & 14 11 16.87 & 39 36 37.06 & $-$910.82 $\pm$ 15.25 & 136.59 $\pm$ 16.18\\
J14122449+1633115 & 14 12 24.50 & 16 33 10.93 & 28.97 $\pm$ 15.90 & $-$80.04 $\pm$ 29.96\\
J14304358+2915405 & 14 30 43.49 & 29 15 41.51 & $-$184.81 $\pm$ 20.15 & 142.49 $\pm$ 16.41\\
J14305589+0013523 & 14 30 55.91 & 00 13 52.03 & 57.81 $\pm$ 24.21 & $-$35.69 $\pm$ 11.98\\
J14321073-0058483 & 14 32 10.72 & 00 58 48.71 & $-$9.00 $\pm$ 20.30 & $-$31.78 $\pm$ 17.14\\
J14385498-1309103 & 14 38 55.08 & $-$13 09 10.57 & 161.23 $\pm$ 21.79 & $-$17.37 $\pm$ 15.99\\
J14393343+0317591 & 14 39 33.43 & 03 17 59.25 & $-$5.58 $\pm$ 20.66 & 19.25 $\pm$ 16.03\\
J14394092+1826369 & 14 39 40.91 & 18 26 36.95 & $-$13.43 $\pm$ 17.75 & 1.26 $\pm$ 25.30\\
J14400180+0021457 & 14 40 01.77 & 00 21 45.90 & $-$73.61 $\pm$ 17.97 & 37.23 $\pm$ 17.96\\
J14413716-0945590 & 14 41 37.07 & $-$09 45 59.20 & $-$203.39 $\pm$ 10.97 & $-$11.16 $\pm$ 13.96\\
J14482563+1031590 & 14 48 25.74 & 10 31 58.34 & 248.93 $\pm$ 15.43 & $-$99.49 $\pm$ 15.58\\
J14493784+2355378 & 14 49 37.86 & 23 55 37.98 & 50.32 $\pm$ 13.25 & 26.69 $\pm$ 23.62\\
J15031961+2525196 & 15 03 19.66 & 25 25 23.56 & 8.98 $\pm$ 17.07 & 20.72 $\pm$ 17.60\\
J15065441+1321060 & 15 06 53.87 & 13 21 06.07 & $-$1086.93 $\pm$ 12.67 & 13.63 $\pm$ 11.38\\
J15150083+4848416 & 15 15 00.24 & 48 47 50.74 & $-$949.93 $\pm$ 21.33 & 1471.48 $\pm$ 21.40\\
J15210327+0131426 & 15 21 03.18 & 01 31 43.12 & $-$211.76 $\pm$ 19.36 & 83.55 $\pm$ 16.58\\
J15394189-0520428 & 15 39 42.18 & $-$05 20 42.02 & 599.49 $\pm$ 13.83 & 117.47 $\pm$ 15.23\\
J15472723+0336361 & 15 47 27.21 & 03 36 36.34 & $-$62.57 $\pm$ 12.75 & 52.34 $\pm$ 16.99\\
J15474719-2423493 & 15 47 47.12 & $-$24 23 50.32 & $-$133.02 $\pm$ 14.53 & $-$121.93 $\pm$ 16.23\\
J15500845+1455180 & 15 50 08.50 & 14 55 17.04 & 105.15 $\pm$ 18.60 & $-$127.21 $\pm$ 12.86\\
15525906+2948485 & 15 52 58.99 & 29 48 48.20 & $-$157.35 $\pm$ 20.31 & $-$50.89 $\pm$ 19.89\\
J15530228+1532369 & 15 53 02.05 & 15 32 38.22 & $-$402.29 $\pm$ 17.20 & 170.52 $\pm$ 16.47\\
J15551573-0956055 & 15 55 16.18 & $-$09 56 22.83 & 928.70 $\pm$ 13.98 & $-$2375.90 $\pm$ 16.67\\
J15552614+0017204 & 15 55 26.04 & 00 17 20.15 & $-$234.00 $\pm$ 17.60 & $-$40.91 $\pm$ 18.14\\
J16000548+1708328 & 16 00 05.47 & 17 08 32.89 & $-$8.80 $\pm$ 15.38 & 15.37 $\pm$ 21.45\\
J16142048+0046434 & 16 14 20.46 & 00 46 43.19 & $-$56.83 $\pm$ 21.23 & $-$31.32 $\pm$ 21.14\\
J16154416+3559005 & 16 15 44.15 & 35 58 56.29 & $-$17.40 $\pm$ 11.94 & $-$512.44 $\pm$ 15.48\\
J16184503-1321297 & 16 18 44.98 & $-$13 21 30.28 & $-$101.25 $\pm$ 14.75 & $-$62.28 $\pm$ 14.63\\
J16192830+0050118 & 16 19 28.33 & 00 50 11.25 & 74.16 $\pm$ 16.12 & $-$88.01 $\pm$ 16.96\\
J16304139+0938446 & 16 30 41.36 & 09 38 44.19 & $-$63.77 $\pm$ 15.27 & $-$55.79 $\pm$ 13.53\\
J16304999+0051010 & 16 30 49.96 & 00 51 00.15 & $-$80.28 $\pm$ 14.87 & $-$138.74 $\pm$ 17.95\\
J16452211-1319516 & 16 45 21.93 & $-$13 19 57.49 & $-$364.32 $\pm$ 17.68 & $-$804.06 $\pm$ 15.75\\
J16573454+1054233 & 16 57 34.51 & 10 54 22.94 & $-$83.53 $\pm$ 17.38 & $-$61.18 $\pm$ 21.24\\
J17054834-0516462 & 17 05 48.40 & $-$05 16 47.06 & 129.32 $\pm$ 14.10 & $-$102.57 $\pm$ 14.92\\
J17073334+4301304 & 17 07 33.21 & 43 01 30.20 &$-$200.33 $\pm$22.02 & $-$21.50$\pm$ 13.67 \\
J17111353+2326333 & 17 11 13.49 & 23 26 32.96 & $-$53.24 $\pm$ 11.33 & $-$35.65 $\pm$ 16.02\\
J17210390+3344160 & 17 21 02.69 & 33 44 20.85 & $-$1853.63 $\pm$ 17.27 & 601.67 $\pm$ 16.84\\
J17260007+1538190 & 17 26 00.05 & 15 38 18.49 & $-$31.20 $\pm$ 12.73 & $-$47.96 $\pm$ 14.38\\
J17282217+5845095 & 17 28 22.19 & 58 45 10.12 & 23.65 $\pm$ 13.39 & 101.69 $\pm$ 12.00\\
J17312974+2721233 & 17 31 29.71 & 27 21 21.79 & $-$81.88 $\pm$ 15.32 & $-$239.88 $\pm$ 17.02\\
J17502385+4222373 & 17 50 23.82 & 42 22 38.02 & $-$43.75 $\pm$ 15.94 & 94.51 $\pm$ 14.29\\
J17580545+4633099&  17 58 05.47 & 46 33 14.57 & 25.51 $\pm$ 15.45 & 593.64 $\pm$ 15.77\\
J18071593+5015316 & 18 07 15.94 & 50 15 30.86 & 34.61 $\pm$ 18.53 & $-$125.71 $\pm$ 14.26\\
J20025073-0521524 & 20 02 50.69 & $-$05 21 53.29 & $-$97.78 $\pm$ 13.02 & $-$104.97 $\pm$ 13.65\\
J20251584-2943124 & 20 26 15.87 & $-$29 43 15.25 & 42.87 $\pm$ 19.44 & $-$347.63 $\pm$ 14.97\\
J20282035+0052265 & 20 28 20.39 & 00 52 26.53 & 114.08 $\pm$ 14.20 & 6.95 $\pm$ 15.27\\
J20343769+0827009 & 20 34 37.66 & 08 26 58.05 & $-$76.65 $\pm$ 15.34 & $-$467.88 $\pm$ 14.90\\
J20360316+1051295 & 20 36 03.11 & 10 51 28.42 & $-$115.22 $\pm$ 13.57 & $-$167.99 $\pm$ 14.07\\
J20543585+1519043 & 20 54 35.83 & 15 19 03.79 & $-$44.56 $\pm$ 17.50 & $-$82.44 $\pm$ 17.60\\
J20571538+1715154 & 20 57 15.42 & 17 15 15.70 & 90.42 $\pm$ 16.27 & 66.20 $\pm$ 14.74\\
J21041491-1037369 & 21 04 15.23 & $-$10 37 39.22 & 614.32 $\pm$ 15.90 & $-$280.70 $\pm$ 14.53\\
J21073169-0307337 & 21 07 31.76 & $-$03 07 33.83 & 169.62 $\pm$ 17.75 & $-$9.94 $\pm$ 13.29\\

\hline
\end{tabular}
\end{center}
\end{table*}
\addtocounter{table}{-1}
 \begin{table*}
\caption{continued}
\begin{center}
\begin{tabular}{l r r r r}
\hline
Name&RA&Dec&$\mu_{\alpha}$cos$\delta$&$\mu_{\delta}$\\
2MASS&\multicolumn {2}{|c|}{J2000}&\multicolumn {2}{|c|}{mas yr$^{-1}$}\\
\hline
J21241387+0059599 & 21 24 13.95 & 01 00 01.49 & 201.78 $\pm$ 14.07 & 287.22 $\pm$ 14.24\\
J21304464-0845205 & 21 30 44.80 & $-$08 45 20.84 & 360.41 $\pm$ 12.75 & $-$31.18 $\pm$ 14.18\\
J21373742+0808463 & 21 37 10.38 & 14 50 46.82 & $-$138.19 $\pm$ 13.82 & $-$122.41 $\pm$ 17.00\\
J21371044+1450475 & 21 37 37.70 & 08 08 46.91 & 704.52 $\pm$ 21.14 & 102.28 $\pm$ 18.58\\
J21404656+0112594 & 21 40 46.51 & 01 12 58.38 & $-$78.42 $\pm$ 20.34 & $-$164.35 $\pm$ 22.03\\
J21522609+0937575 & 21 52 26.20 & 09 37 58.42 & 293.64 $\pm$ 18.65 & 170.15 $\pm$ 17.32\\
J21580457-1550098 & 21 58 04.60 & $-$15 50 10.14 & 69.64 $\pm$ 11.27 & $-$33.12 $\pm$ 14.92\\
J22081363+2921215 & 22 08 13.70 & 29 21 21.33 & 111.25 $\pm$ 14.37 & $-$11.47 $\pm$ 14.05\\
J22134491-2136079 & 22 13 44.94 & $-$21 36 08.56 & 59.56 $\pm$ 10.65 & $-$62.73 $\pm$ 16.88\\
J22425317+2542573 & 22 42 53.40 & 25 42 56.94 & 408.77 $\pm$ 15.46 & $-$45.19 $\pm$ 16.17\\
J22443167+2043433 & 22 44 31.81 & 20 43 41.34 & 251.74 $\pm$ 13.97 & $-$213.74 $\pm$ 10.81\\
J22521073-1730134 & 22 52 10.92 & $-$17 30 12.48 & 404.85 $\pm$ 20.23 & 153.57 $\pm$ 20.33\\
J22541892+3123948 & 22 54 18.96 & 31 23 51.33 & 67.94 $\pm$ 15.11 & 200.74 $\pm$ 11.26\\
J22545194-2840253 & 22 54 51.94 & $-$28 40 25.20 & 8.20 $\pm$ 18.69 & 51.81 $\pm$ 30.09\\
J22591388-0051581 & 22 59 13.91 & 00 51 57.87 & 83.81 $\pm$ 16.71 & 67.87 $\pm$ 16.80\\
J233302258-0347189 & 23 30 22.71 & $-$3 47 18.84 & 231.86 $\pm$ 17.43 & 31.93 $\pm$ 13.63\\
J23440624-0733282 & 23 44 06.25 & $-$7 33 28.72 & 21.34 $\pm$ 16.63 & $-$51.27 $\pm$ 7.71\\
J23453909+0055137 & 23 45 39.07 & 00 55 13.45 & 100.97 $\pm$ 18.74 & $-$34.01 $\pm$ 14.45\\
J23515044-2537367 & 23 51 50.65 & $-$25 37 35.27 & 377.56 $\pm$ 20.85 & 210.93 $\pm$ 22.14\\
\hline
\end{tabular}
\end{center}
\end{table*}

Since the observations were obtained, a paper by \citet{schmidt07} has 
been published. This paper contains proper motions for 23 of these 143 objects. The proper motions as calculated here and as presented in \citet{schmidt07} are shown in table \ref{previous}.
\begin{table*}
\caption{\label{previous}Name,  proper motion (from \citet{schmidt07}), position angle (from \citet{schmidt07}),
proper motion (from this study), position angle ( from this study) for the 23 dwarfs  which appear in \citet{schmidt07}}
\begin{center}
\begin{tabular}{l r r r r}
\hline
Name&$\mu$ (Schmidt)&PA (Schmidt)&$\mu$&PA\\
2MASS&'' yr$^{-1}$&$^{\circ}$&'' yr$^{-1}$&$^{\circ}$\\
\hline
J0045+16&0.36$\pm$0.04&102$\pm$7&0.38$\pm$ 0.02&93$\pm$2\\
J0251-03&2.17$\pm$0.11&148$\pm$2&2.14$\pm$0.02&148$\pm$0.6\\
J0908+50&0.52$\pm$0.25&202$\pm$21&0.61$\pm$ 0.02&220$\pm$1\\
J1045-02&0.46$\pm$0.12&266$\pm$8&0.49$\pm$0.02&272$\pm$1.5\\
J1048+01&0.52$\pm$0.1&241$\pm$9&0.48$\pm$0.01&240$\pm$0.6\\
J1051+56&0.32$\pm$0.09&197$\pm$23&0.36$\pm$0.0&219$\pm$3\\
J1104+19&0.14$\pm$0.08&24$\pm$31&0.15$\pm$0.02&28$\pm$1\\
J1203+00&1.00$\pm$0.15&257$\pm$6&1.23$\pm$0.02&258$\pm$0.5\\
J1213-04&0.15$\pm$ 0.09&260$\pm$38&0.35$\pm$0.02 &268$\pm$2\\
J1221+02&0.09$\pm$ 0.16 &258$\pm$76&0.11$\pm$0.02&262$\pm$11.5\\
J1448+10&0.71$\pm$ 0.15&98$\pm$13&0.26$\pm$0.02&111$\pm$4\\
J1506+13&1.11$\pm$ 0.08&274$\pm$3&1.08$\pm$0.01&271$\pm$0.6\\
J1515+48&1.71$\pm$ 0.05& 327$\pm$1&1.75$\pm$0.02&328$\pm$1\\
J1539-05&0.6$\pm$ 0.04&85$\pm$7&0.61$\pm$0.01&78$\pm$1\\
J1721+33&1.92$\pm$ 0.11& 287$\pm$3&1.94$\pm$0.02&288$\pm$0.6\\
J1731+27&0.28$\pm$ 0.05&200$\pm$5&0.25$\pm$0.02&199$\pm$2\\
J1807+50&0.16$\pm$ 0.05&167$\pm$15&0.13$\pm$0.01&164$\pm$9\\
J2028+00&0.14$\pm$ 0.09&105$\pm$28&0.11$\pm$0.01&86$\pm$7\\
J2036+10&0.26$\pm$ 0.09&216$\pm$21&0.20$\pm$0.01&215$\pm$1\\
J2057-02&0.08$\pm$ 0.02&171$\pm$17&0.07$\pm$0.01&170$\pm$12\\
J2104-10&0.66$\pm$ 0.05&115$\pm$3&0.68$\pm$0.02&114$\pm$1.6\\
J2251-17&0.45$\pm$ 0.12&71$\pm$10&0.43$\pm$0.02&69$\pm$1.6\\
J2351-25&0.42$\pm$ 0.11&62$\pm$15&0.43$\pm$ 0.02 &60$\pm$01\\
\hline
\end{tabular}
\end{center}
\end{table*}

There are two objects for which we do not find a good agreement between this data and the results of \citet{schmidt07}. These objects are J1213$-$04 and J1448+10.
Both J1213$-$04 and J1448+10 have position angles that agree, but the proper motion measurements do not. The discrepancy involving J1213-04 can almost be explained by the errors on the \citet{schmidt07} measurement ($\mu_{\alpha}$=144$\pm$80, $\mu_{\delta}$=-25$\pm$98). Our $\mu_{\delta}$ agrees with that of \citet{schmidt07}, although the large disagreement with $\mu_{\alpha}$ (our measurement=248$\pm$15, Schmidt et al. measurement=708$\pm$142) may be explained by the object being faint and a small epoch difference ($<$4 years) for the \citet{schmidt07} measurement.

There are two notes that should be made regarding these 143 objects in table \ref{pm}. The first is that J1048+01 has a somewhat unclear spectral type. Using an optical 
spectrum, \citet{hawley02} identify it as an L1 dwarf. Using infrared spectra, \citet{kendall04} identify it as an L4 dwarf, and \citet{wilson03} identify it as an M7 
dwarf also using infrared spectra. It is unknown as to why this discrepancy has occurred. 

The second note is that J2213$-$21 has been identified as a possible low gravity object by \citet{cruz07} from the VO bands, K I doublet and Na I doublet in the optical spectrum of this object. These features are gravity sensitive, and low gravity features indicate youth and low mass. H$\alpha$, also an indicator of youth and activity was not detected in the spectrum.
\section{Common Proper motion objects}

Brown dwarfs in binary or multiple star systems are of great interest as their properties such as masses and separations can allow constraints to be put on star formation models. This is especially important as no one mechanism has been found that can account for the formation of all known brown dwarf binaries. Many imaging surveys have been undertaken to find close brown dwarf binaries however very few wide brown dwarf binaries are known.  Only one brown dwarf-brown dwarf system has been discovered to date with a separation of greater than 15 AU \citep{billeres05}, in contrast to several known systems with a high mass primary and brown dwarf companion. \citet{allen07} estimate the wide binary fraction for these objects to be 2.3\% for companions with masses greater than 0.03-0.05 M$_{\odot}$.

We selected a ``standard'' $\sigma$ value of 15 mas yr$^{-1}$, within which to search for common proper motion companions to the L and T dwarfs.
This value was chosen as an average value of the errors, and therefore this search is by no means conclusive. 
However, as some objects have very little proper motion compared to the background stars, searching for common proper motion objects yields many candidates. 
No more can be said about these objects, other than they are candidates and more information is required.
For the objects that have proper motions that are much higher than the background stars (129 out of the 143 objects), as we would expect for nearby L and T dwarfs, searching for proper motion companions (i.e. objects with proper motions that fall within a radius of 15 mas yr$^{-1}$ of the motion of the dwarf) is more robust, and yields a few results.
We find;\\
\begin{enumerate}
\item Proper motions for 6 previously known close binaries, J1239+55, J1430+29, J1553+15, J1600+17, J2152+09 and J2252$-$17.\\
\item We confirm two previously known common proper motion wide binaries. J1004+50 also known as Gl96-3B and its companion Gl96-3A \citep{salim03}. J1441$-$09 is a binary consisting of two L1 dwarfs with an M4.5 primary \citep{seifahrt05}.\\
\item We find 6 new possible multiple systems, although one appears unlikely, see table \ref{bin} and comments below.\\
\end{enumerate}

Of the 6 new systems,  4  have not previously been identified as binaries. They are J1153+50, J0041+13, J1547+03 and J2259$-$00. These objects all appear to have proper motion companions. The proper motions, distance from the dwarf, \textit{J},\textit{H} and \textit{K$_{S}$} magnitudes of these objects and the L dwarf ``primary'' are given in table \ref{bin}. J1017+13 and J144923 are known binaries to which we add another wide proper motion component.\\

\begin{table*}
\caption{\label{bin}Name, proper motion for each component, \textit{J}, \textit{H} and \textit{K$_{S}$} magnitudes and distance from L dwarf for the 6 new possible binaries.}
\begin{center}
\begin{tabular}{l c c r r r r}
\hline
Name&$\mu_{\alpha}$cos$\delta$&$\mu_{\delta}$&\textit{J}&\textit{H}&\textit{K$_{S}$}&Distance\\
&\multicolumn {2}{|c|}{mas yr$^{-1}$}&&&&''\\
\hline
J0041+13&$-$174.28$\pm$23.60&$-$137.55$\pm$36.02&14.454$\pm$0.031&13.673$\pm$0.036&13.236$\pm$0.024&$-$\\
2MASSJ00415543+1341162&$-$182.28$\pm$23.60&$-$141.99$\pm$36.02&10.164$\pm$0.029&9.574$\pm$0.033&9.347$\pm$0.018&32.02\\
\hline
J1017+13&60.53$\pm$21.83&$-$93.74$\pm$12.29&14.096$\pm$0.021&13.284$\pm$0.026&12.710$\pm$0.021&$-$\\
2MASSJ10171515+1307419&58.08$\pm$21.83&$-$93.74$\pm$12.29&16.827$\pm$0.157&16.497$\pm$0.296&17.365&125.32\\
\hline
J1153+50 &83.92 $\pm$22.95 & 60.19$\pm$10.88& 14.189$\pm$0.026&13.305$\pm$0.024&12.851$\pm$0.025&$-$\\
2MASSJ11535018+5035593&87.50$\pm$22.95&70.37$\pm$10.88&17.145$\pm$0.173&16.349$\pm$0.233&15.832$\pm$0.203&251.12\\
\hline
J1449+23&40.31$\pm$13.25&26.69$\pm$23.61&15.818$\pm$0.074&15.004$\pm$0.086&14.311$\pm$0.084&$-$\\
2MASSJ14493550+2357118&48.3$\pm$13.25&30.71$\pm$23.61&16.663$\pm$0.154&17.709&14.311&99.36\\
\hline
J1547+03&$-$62.56$\pm$12.74 &52.33$\pm$16.98&16.077$\pm$0.071&15.070$\pm$0.062&14.270$\pm$0.072&$-$\\
2MASSJ15470234+0338260&$-$51.91$\pm$12.74&57.67$\pm$16.98& 16.275$\pm$0.094&15.843$\pm$0.124&15.691$\pm$0.233&388.44\\
\hline
J2259$-$00&83.81$\pm$16.70&67.87$\pm$16.79&16.357$\pm$0.097&15.315$\pm$0.087&14.651$\pm$0.087&$-$\\
2MASSJ22590929$-$0051556&69.93$\pm$16.70&72.91$\pm$16.79&11.264$\pm$0.026&10.993$\pm$0.026&10.872$\pm$0.021&68.80\\
2MASSJ22590491$-$0052407&87.59$\pm$16.70&60.30$\pm$16.79&17.123$\pm$0.214&16.300$\pm$0.231&15.692$\pm$0.229&141.11\\
\hline
\end{tabular}
\end{center}
\end{table*}

\textbf{J0041+13} has a spectral type of L0 \citep{hawley02} and has very similar proper motion to a brighter field object
2MASSJ00415543+1341162 is also known as NLTT 2274, and is labelled as a high proper motion star.SIMBAD, lists  a proper motion of $-$191,$-$167 mas yr$^{-1}$
\citep{salim03}, which is consistent with our measurements.\\

\textbf{J1017+13}, a brown dwarf of spectral type L2 \citep{cruz03} was reported as a candidate binary system by \citet{bouy03}. Their HST imaging program using WFPC2 determined that  it has a separation of
 104$\pm$2.8 mas and a position angle of 92$^{\circ}$.6$\pm$1$^{\circ}$.2. The two components are found to have differences in magnitudes, suggesting they 
are of different mass.
The photometric distance for this object was calculated to be $\approx$21.4 pc. In our data this dwarf appears to have common motion with another object in the field,
 2MASSJ10171515+1307419.  This wide companion appears to be a T dwarf at the distance of 22.8 pc (as calculated from spectral type of the L dwarf)
from its colours \citep{leggett02} and absolute magnitude. This object does form a sequence with the L2 primary, which may also be a binary. 
We caution that this object is faint so the 2MASS measurement and the 
WFCAM measurement of it are of low S/N. More information is needed before conclusions can be drawn.\\

\textbf{J1153+50} has a spectral type of L1 (Reid et al., in preparation) and appears to have a much fainter field object associated with it. Assuming a distance of 27.4 pc for this dwarf (calculated from its spectral type using equation \ref{1}) and the companion, the colours of the companion indicate a spectral type of between L5 and L8 \citep{leggett02} and the absolute magnitude of the companion fit the L-T dwarf sequence, strengthening the case that these dwarfs are companions.\\

\textbf{J1449+23} is reported as being a binary by \citet{gizis03}. This is currently one of the widest L dwarf doubles known, at a separation of 13 AU. J1449+23 has a spectral type of L0 \citep{kirkpatrick00} and it is not known if the components are very low mass stars or brown dwarfs.
This pair appears to have very similar proper motion to another object in the field,
 J14493550+2357118. 
J14493550+2357118 has impossibly blue colours ($J$-$K$=$-$0.197, $J$-$H$=$-$1.046), but it is at the survey limit of 2MASS, making it likely these measurements have large errors. In fact this object is so faint that errors are not given for some of the 2MASS photometry. It may thus be considered that this object is only detected in the $J$ band. If the $J$, $H$ and $K$ photometry proves to be accurate then it is possible that this object is a white dwarf. When the SDSS colours are compared to the \citet{holberg06} synthetic hydrogen white dwarf photometry, it falls along the hydrogen track, with a mass of 0.2 M$_{\odot}$, T$_{\rm eff}$$\approx$4000, log g$\approx$1.75 if the optical Sloan colours are used. 
A spectrum is needed to substantiate this hypothesis.\\

\textbf{J1547+03} has a spectral type of L2 \citep{hawley02}  and appears to be associated with a field object, that appears to be only slightly fainter. This object however has blue $J$-$K$ and $J$-$H$ colours that are not consistent with a T dwarf of the same distance as J1547+03 as it would be too bright \citep{leggett02}. The Sloan photometry and JHK photometry is however
consistent with a white dwarf at the same distance (58.8 pc) as the brown dwarf according to the Helium track of the \citet{holberg06} models. This white dwarf would have a log g of 7.4, a T$_{\rm eff}$=5500 and a mass of 0.3M$_{\odot}$. This white dwarf would have an age of $\approx$ 1.7 Gyr. If this object is a white dwarf it will be one of only a few confirmed white dwarf$-$brown dwarf known binaries (GD165 \citet{becklin88}, GD1400 \citet{farihi04}, WD0137-349 \citet{burleigh06a}, SDSS1212 \citet{burleigh06b} although others are suspected.
However, this white dwarf may also be a background high proper motion, metal poor K dwarf, which would also exhibit these colours. Many common proper motion
objects are in fact high velocity background dwarfs, particularly at large separations \citep{farihi05}. A spectrum of this object was taken using ISIS on the William Herschel Telescope on 18 August 2007. Spectra were taken in the red arm using the R316 grating and the R300 grating in the blue arm. This spectrum proved this ``companion'' to be a metal poor K dwarf. \\

\textbf{J2259-00} has a spectral type of L2 \citep{hawley02} and has similar proper motion to two field objects, one which is much brighter and one, which is much fainter.
This fainter object has a possible spectral type of L3 (using the colours and \citet{leggett02}), at a distance of 64.7 (derived from spectral type). However, when plotted on the M$_{\rm K}$, \textit{J}-\textit{K} 
colour magnitude diagram, these three objects do not form a sequence. The brighter object has colours that are approximately those of a G8 dwarf main sequence star, 
but at this distance, the object in question is too faint. We do not believe that this brighter object is a companion to the dwarf. The fainter companion also does not lie on a sequence with the L2 dwarf. We believe that these three objects have similar proper motions, but are not related.\\

Out of the 129 dwarfs with isolated enough motion to study, 5 appear to have companions. Of these 5, only 2 appear to be possible brown dwarfs. From this study the wide brown dwarf-brown dwarf binary fraction is 1.55\%, which is in agreement with the 2.3\% upper limit calculated by \citet{allen07}. 
This fraction will not include any binaries with very faint companions due to the limiting depth of 2MASS. Many more such binaries will ultimately be found by the UKIDSS large area survey when its second epoch for proper motions is complete.

\section{High Velocity L dwarfs}
\begin{figure}
\begin{center}
\scalebox{1.20}{\includegraphics[angle=270, width=\linewidth]{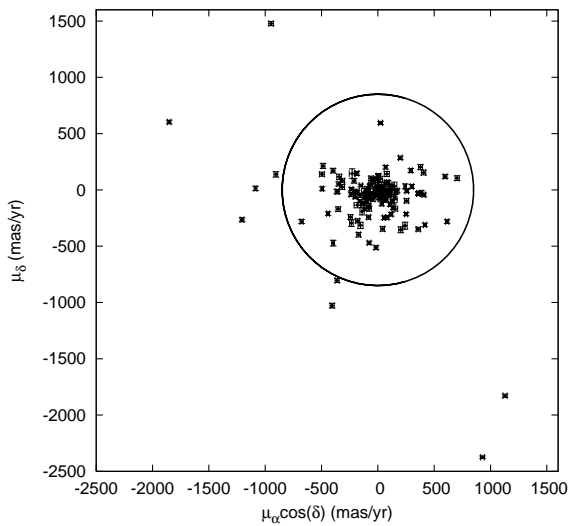}}
\caption{
\label{pm_plot}
Proper motions of the 143 brown dwarfs. The circle has a radius of 0.85''. The objects outside this circle, are the nine fast moving objects.}
\end{center}
\end{figure}

From the proper motions calculated using the WFCAM and 2MASS data,we selected all (nine in total) objects with an extremely high proper motion ($>$0.85'' yr$^{-1}$, figure \ref{pm_plot}).
Five of these nine objects appear in \citet{schmidt07} and two are mentioned in the text as having large proper motions. Our proper motion and position
 angle measurements agree with those that appear in table 1 of \citet{schmidt07} to within the errors  as shown in table \ref{highvel}.
Tables \ref{highvel} and \ref{highvelocity} show the data for the  L dwarfs selected as high velocity L dwarfs.  The distances were calculated using the M$_{J}$ spectral type relation given by \citet{cruz03}, equation \ref{1}. 
\begin{eqnarray}
\label{1}
M_{J}&=&-4.410+5.043(ST)-0.6193(ST)^2+ \nonumber \\
&&0.03453(ST)^3-0.0006892(ST)^4,
\end{eqnarray} 

where ST=0, 10 and 18 for spectral types M0, L0 and L8 respectively.
\begin{table*}
\caption{\label{highvel}Name, spectral type( from optical spectra), distance (as calculated from spectral type), proper motion (from \citet{schmidt07}), position angle(from \citet{schmidt07}),
proper motion (from this study), position angle ( from this study) for the 8 high velocity dwarfs, 5 of which appear in \citet{schmidt07}.}
\begin{center}
\begin{tabular}{l c r r r r r}
\hline
Name&Spt& d&$\mu$ (Schmidt)&PA (Schmidt)&$\mu$&PA\\
&&pc&'' yr$^{-1}$&$^{\circ}$&'' yr$^{-1}$&$^{\circ}$\\
\hline
J0251-03&L3&11.98&2.17$\pm$0.11&148$\pm$2&2.14$\pm$0.02&148$\pm$0.6\\
J1203+00&L3&18.53&1.00$\pm$0.15&257$\pm$6&1.23$\pm$0.02&258$\pm$0.5\\
J1331-01&L6&19.53&$-$&$-$&1.07$\pm$0.01&202$\pm$0.6\\
J1411+39&L1.5&31.56&$-$&$-$&0.92$\pm$0.02&279$\pm$01.1\\
J1506+13&L3&13.80&1.11$\pm$0.08&274$\pm$3&1.08$\pm$0.01&271$\pm$0.6\\
J1515+48&L6&10.50&1.71$\pm$0.05&327$\pm$1&1.75$\pm$0.02&328$\pm$1\\
J1555-09&L1&12.94&$-$&$-$&2.55$\pm$0.02&158$\pm$0.5\\
J1645-13&L1.5&11.52&$-$&$-$&0.88$\pm$0.01& 205$\pm$1\\
J1721+33&L3&15.56&1.92$\pm$0.11&287$\pm$3&1.94$\pm$0.02&288$\pm$0.5\\
\hline
\end{tabular}
\end{center}
\end{table*}
J1645$-$31, while having a relatively high proper motion, is close and has a small tangential velocity, and so is not believed to actually be be a fast moving object.

We have calculated an approximate error of 13\% on the distance, based on the scatter in the Cruz relation, and an assumed uncertainty of $\pm$ half a spectral class. \citet{cruz03} state that the error on this relationship comes from the uncertainties in spectral types. This is a tighter relationship than using \textit{J}-\textit{K$_{S}$} colours to derive M$_{J}$, which shows a fair amount of scatter. The tangential velocities were then be derived from the 
proper motions and distances. Most of the eight dwarfs have tangential velocities exceeding 70 km s$^{-1}$. The error on the velocities has been taken to be 13\%, dominated by the 
distance error. \citet{schmidt07} find the mean tangential velocity for a sample of M and L dwarfs to be 31.5 km s$^{-1}$, with a velocity dispersion of 20 km s$^{-1}$. Thus these L dwarfs clearly have extreme velocities and the natural interpretation is that they belong to the old disc, thick disc or possibly galactic halo 
population. \citet{schmidt07} also identified two of these dwarfs, J1721+33 and J2051-03 as high velocity dwarfs.

To better understand the motion of these fast moving objects, we transformed the proper motion measurements in right ascension and declination ($\mu_{\alpha}cos\delta$, $\mu_{\delta}$) into proper motion on the galactic co$-$ordinate ($\mu_{l}$, $\mu_{b}$) system and calculated the velocities.
To be able to study these velocities in relation to the galaxy, these velocities were then transformed onto a 
right handed co$-$ordinate system, with orthogonal axes directed towards the Galactic centre, 
the U velocity, in the direction of Galactic rotation (l=90$^{\circ}$, b=0$^{\circ}$), V velocity and perpendicular to the galactic plane, 
W velocity. The radial velocity is unknown for these dwarfs, however if l$\approx$0 or 180$^{\circ}$ it is possible to find V, since V$_{r}$sinl$\approx$0 
Likewise if l$\approx$90 or 270$^{\circ}$ it is possible to find U.

Table \ref{highvelocity} gives approximate U and V velocities for those dwarfs close to such special locations. 
J0251-03 and J1555-09  have V=$-$120.20 km s$^{-1}$ and V=-73.18 km s$^{-1}$  respectively and J1203+00 and J1515$-$09 have U=$-$81.61 and U=$-$86.81 km s$^{-1}$ respectively.
These velocities  are typical of a thick disc population (Burgasser, Cruz \& Kirkpatrick, 2007a).
All eight L dwarfs have spectral types derived from optical spectra \citep{cruz03, fan00, hawley02, kirkpatrick01, gizis00, wilson03,gizis02} and none of the authors comment on any of the spectra being abnormal.

\begin{figure*}
\begin{center}
\scalebox{0.5}{\includegraphics[angle=270]{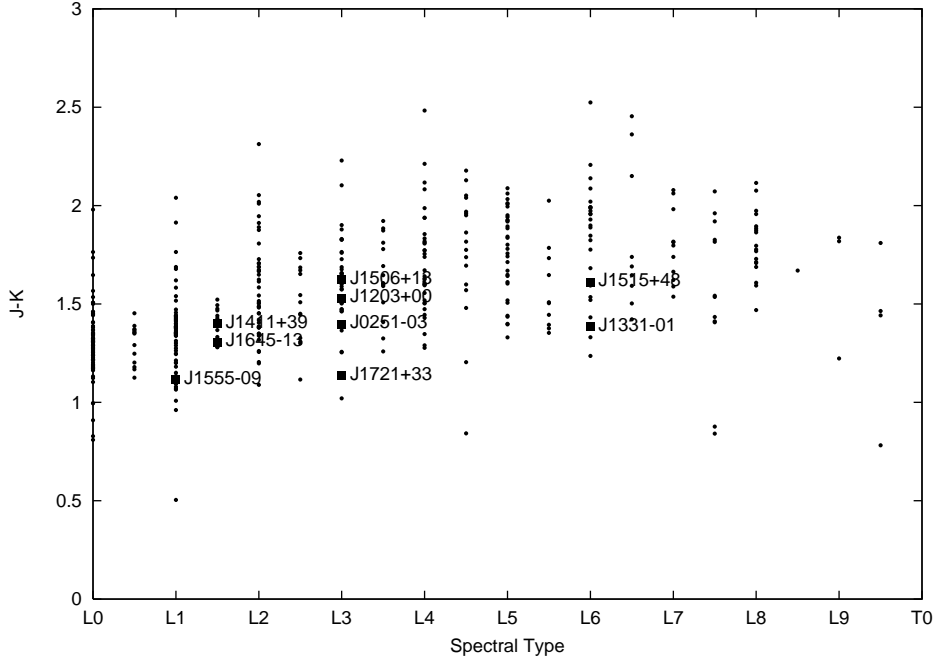}}
\caption{\label{spt}\textit{J}-\textit{K$_{S}$} vs spectral type for all L dwarfs in the Dwarf Archive. The marked objects are the 9 fast moving dwarfs which tend to have bluer \textit{J}-\textit{K$_{S}$} than other dwarfs of the same spectral type.}
\end{center}
\end{figure*}

\citet{burgasser07a} discuss 16 spectroscopically confirmed subdwarfs. The L4 subdwarf 2MASS J1626+3925 shows a broad dip from 6700 - 7300 \AA \space when compared to a ``standard'' L4 dwarf. This dip is much less pronounced in the L7 subdwarf 2MASS J0532+8246 when compared to a ``standard'' L7 dwarf. Indeed, it is difficult to 
distinguish the L7 subdwarf from an ordinary L7 dwarf by its 6500 - 9000 \AA \space spectrum. 
The 6700 - 7300 \AA \space dip is not obvious in the published spectrum of the L3 dwarf J1721+33 and is not expected in the L6 spectra of J1331-01 and J1555-09.
Thus the only supporting evidence that these high velocity dwarfs are subdwarfs is their blue $J$-$K_{S}$ colour. Blue $J$, $H$ and $K_{S}$ colours are expected for metal poor stars since the collisionally induced molecular hydrogen opacity is relatively more important and this increases rapidly from $J$ to $K$ depressing the longer wavelengths \citep{burgasser07}. Figure \ref{spt} shows $J$-$K_{S}$ for all the L dwarfs in the dwarf archive. Our 8 high velocity dwarfs are marked as squares followed by their names. It can be seen that they all have bluer than average $J$-$K$ colours. These colours however are not blue enough to indicate that these objects are subdwarfs as defined by \citep{burgasser07a}.

If we  assume that all 8 of these dwarfs are likely to belong to the thick disc population (they are not blue enough to belong to the population II halo) then it is to be expected that they are old, with an age of $\approx$10 Gyr.
This is most likely for the three having the most obviously blue  \textit{J}-\textit{K$_{S}$} colour, J1721+33, J1331-01 and J1555$-$09. With an age of $\approx$10 Gyr however they cannot be brown dwarfs, which would have cooled to much lower effective temperatures \citep{burrows01}. 
We suspect that  they are in fact extremely low luminosity stars. To check this, we have calculated their luminosities. \citet{leggett02} find that the \textit{K$_{S}$} band bolometric correction is 3.3 for L dwarfs. Using this number and the absolute \textit{K} magnitude we can calculate L/L$_{\odot}$ as shown in table \ref{highvelocity}.
It should be noted that assuming BC$_{\rm K}$ =3.3 may be slightly unreliable for dwarfs that are unusually blue in \textit{J}-\textit{K}.
These  L/L$_{\odot}$ $\approx$ 10$^{-4}$ are in agreement with theoretical models for very low mass old stars \citep{burrows01}. Indeed, J1331$-$01, J1515+48 may be the lowest luminosity stars found to date.

\begin{table*}
\caption{\label{highvelocity}Name, l, b, V$_{\rm Total}$ (total velocity), U( velocity in the direction of the galactic centre), V (velocity in the direction of galactic rotation), \textit{K}, \textit{J}-\textit{K} and L/L$_{\odot}$for the eight fast moving dwarfs.}
\begin{center}
\begin{tabular}{l c c c c c c c c}
\hline
Name&l&b&V$_{\rm Total}$&U&V&\textit{K}&\textit{J}-\textit{K}&L/L$_{\odot}$\\
&\multicolumn {2}{|c|}{$^{\circ}$}& \multicolumn {3}{|c|}{km s$^{-1}$}&&&$\times$10$^{-4}$\\
\hline
J0251$-$03&179.10& $-$53.15&121.97& $-$&   $-$120.20    &11.662 &1.39&1.18\\
J1203+00&277.98& 60.81&108.71& $-$81.61&    $-$  &12.476 &1.53&1.33\\
J1331$-$01& 323.42&  59.97& 102.55&  $-$& $-$ &14.073&1.38&0.34\\
J1411+39&74.98& 69.25& 137.81&$-$ &   $-$   &13.239 &1.40&1.91\\
J1506+13&16.21& 55.52&  71.10&$-$  &  $-$    &11.741 &1.62&1.45\\
J1515+48&80.75& 54.88 &87.18&$-$86.81&     $-$   &12.500 &1.61&0.41\\
J1555$-$09&359.58&  32.05&156.52& $-$&    $-$73.18   &11.443&1.14&1.68\\
J1721+33&57.35 &  32.52 &143.70&$-$    &$-$&12.489 &1.36&0.92\\
\hline
\end{tabular}
\end{center}
\end{table*}

\section{Moving groups}
Following the method of \citet{bannister07}, we used the moving cluster method to search for members of the Ursa Major/Sirius, Hyades and Pleiades moving groups.  If any of the L and T dwarfs that we have measured proper motions  appear to be moving towards the convergent point of any of these groups, then it is a potential member. The moving group distance can then be calculated from equation \ref{2}, 
\begin{equation}
\label{2}
d_{mg}=\frac{v\sin\theta}{4.74\mu},
\end{equation}

where v is the moving group velocity (km s$^{-1}$), $\theta$ the angular distance of the dwarf from the convergent point, and $\mu$ the proper motion in arcseconds.

\citet{bannister07} used objects with known parallax measurements, to check if the distance provided by the moving cluster method was comparable to the measured distance. We do not have parallax measurements for any of our objects, and so must rely on a different distance check. We have again used the Cruz distance relationship  based on spectral type (equation \ref{1}).
Once the  distance to the star, assuming it is a moving group member has been calculated, the distance formula  can be used to 
calculate the absolute magnitude of the object in question, and plot it on a colour$-$magnitude diagram.
The objects that have been selected - i.e. have the correct proper motion angle, which must be less than 14$^{\circ}$ for the difference between the observed and calculated angles as in \citet{bannister07}, and  the correct distance to fit on a colour-magnitude diagram, were
then subject to another check.  

If the ratio between the 
moving cluster distance and the Cruz spectral type distance was greater than 0.72 and less than 1.28, as calculated in \citet{bannister07} then the objects were considered to be members. Parallax measurements for these members could confirm these distances.
While some objects may have the correct direction and magnitude of motion, few will have the correct distances and appear to sit in the correct position on the colour magnitude diagram. In their study, \citet{bannister07} estimated 0.5\% probablility of an object passing all three tests (distance, direction and colour magnitude diagram). They suggest that it is likely that 99.5\% of selected objects therefore are members of the moving groups. 

In our case we only have spectroscopic distance estimates to compare with the moving group distance, so the group membership probabilities will be lower.
The random chance probability of passing the direction test is naively 2$\times$14/360=0.08. However, this assumes that field dwarf proper motions are randomly orientated.
Figure \ref{pm_plot} shows that they are approximately orientated in a semicircle, so we therefore double the random chance to 0.16. Using a spectroscopic distance of course results in a reasonable colour-magnitude diagram i.e. one that is appropriate to the mean dwarf age, so the third test presented in \citet{bannister07} is not valid.
However, requiring that the moving group distance is approximately equal to the spectroscopic distance is still a valid test since it effectively requires the candidate to have the ``correct'' amplitude of proper motion.
The sample of 143 proper motions peaks at an amplitude of 90 mas yr$^{-1}$. This number then falls off exponentially towards high proper motions. However, very few of the moving group candidates have proper motions $>$200 mas yr$^{-1}$. Thus we take the width of the proper motion distribution to be 30 to 200  mas yr$^{-1}$, a factor of 200/30=6.67. The errors in the spectroscopic distance is taken to be 13\% (see above) and we allow a 28\% varaition in d$_{mg}$/d$_{sp}$ due to the moving group velocity dispersion.
Combining these percentages quadratically gives  $\pm$31\% i.e. a distance factor 1.31/0.69=1.9.  Now, 6.67$\sim$(1.9)$^{3}$ so there is a 1 in 3 random chance of the moving group distance agreeing with the spectroscopic distance.
Thus the total random chance of membership is 0.16$\times$0.33=0.05. These rather crude statistics therefore suggest that about 95\% of our moving group candidates should be genuine members.
It should be noted however that galactic resonances can also produce common velocities similar to those of a moving group \citep{dehnen98}.
Thus members of a moving group may have the same age, but coevality is not guaranteed.

\subsection{The Hyades}

The Hyades cluster has a distance of 46 pc and covers $\approx$ 20$^{\circ}$ of the sky. The Hyades has almost no known low mass members \citep{gizis99, dobbie02} as due to its age (625 Myr \citet{perryman98}) they have  evaporated from the cluster.In fact for a cluster of this age $\sim$ 70\% of stars and $\sim$ 85\% of brown dwarfs are expected to have escaped the cluster \citep{adams02}. The Hyades moving group is made up of objects that may have escaped from the Hyades. \citet{chereul98} first identified escaped Hyads, and more recently \citet{bannister07} have identified 7 L and T field dwarfs that belong to the Hyades moving group. \citet{zapatero07} have confirmed membership for one of these objects (2MASS J1217110-031113), and have disproved membership for two (2MASS J0205293-115930 and 2MASS J16241436+0029158) using radial velocity measurements. The Hyades moving group has its convergent point situated at $\alpha$=6$^{h}$29.48$^{m}$, $\delta$=-6$^{\circ}$53.4', and the members have a space velocity of 46 km s$^{-1}$ (Madsen, Dravins \& Lindgren, 2002). 

\begin{figure*}
\begin{center}
\scalebox{0.50}{\includegraphics[angle=270]{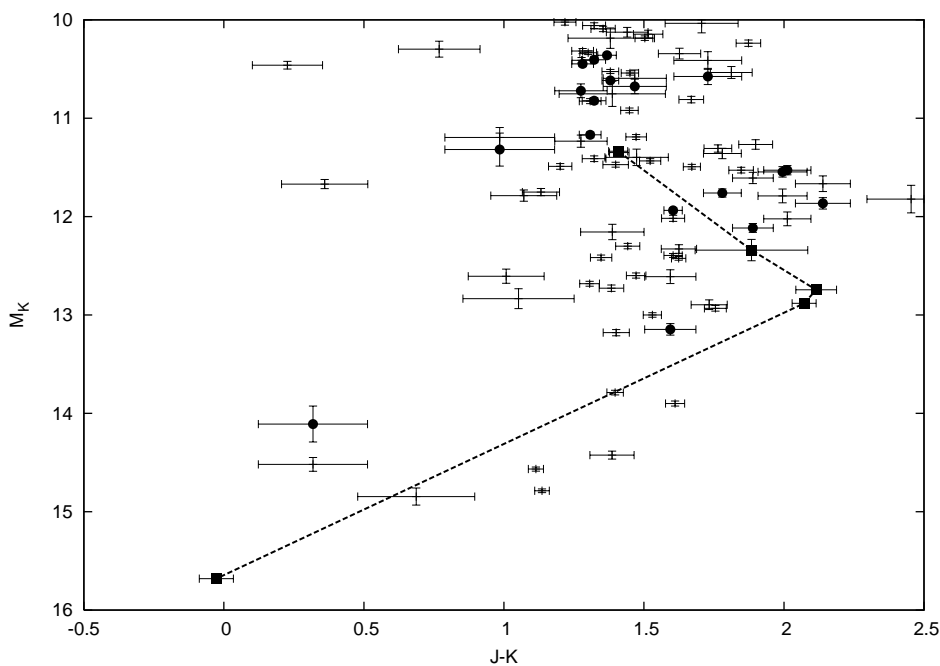}}
\caption{
\label{mg_hyadesall}
M$_{\rm K}$, $J$-$K$ colour magnitude diagram for the Hyades moving group. The members identified by \citet{bannister07} are plotted as filled squares. All of the 143 objects are plotted, with M$_{\rm K}$ calculated from the moving group distance. The objects selected from the moving group method (correct angle of proper motion and distance ratio) are marked as filled circles. The errors are poissonion and from the photometry only. The dashed line indicates the possible single star sequence as calculated from the Bannister objects.}
\end{center}
\end{figure*}

\begin{figure*}
\begin{center}
\scalebox{0.50}{\includegraphics[angle=270]{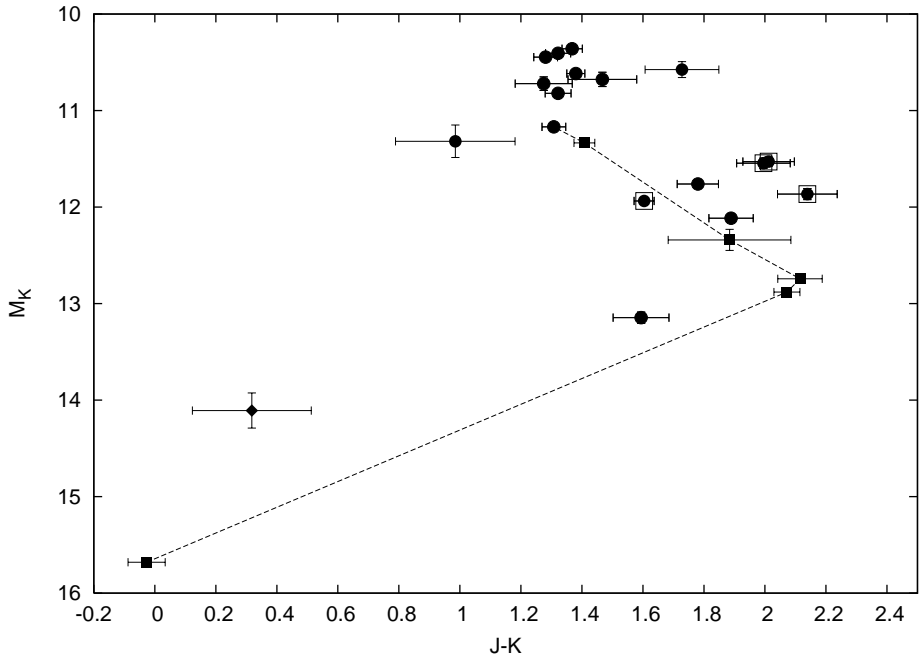}}
\caption{
\label{mg_hyades}
M$_{\rm K}$, $J$-$K$ colour magnitude diagram for the Hyades moving group. The members identified by \citet{bannister07} are plotted as filled squares.  All of the objects that were selected as possible members are marked as filled circles and the selected members are marked by a ring around the filled circle. The diamond is the object J1553+15.The 4 objects that are uncertain members are marked with boxes around the filled circles.The errors are poissonion and from the photometry only.The dashed line indicates the possible single star sequence.}
\end{center}
\end{figure*}

Fifteen objects had the correct proper motion, direction (distance between the observed angle of proper motion and moving group angle must be less than 14$^{\circ}$) and distance (ratio between the 
moving cluster distance and the Cruz spectral type distance  greater than 0.72 and less than 1.28) to belong to the Hyades moving group as shown on figure \ref{mg_hyadesall}. 
Of these, 8 are possible binaries as they lie up to 0.75 magnitudes above the main sequence.
J1047$-$18 and J1750+42 had the correct motion, direction and distance to belong to the moving group, however lay in the wrong place on the sequence in the M$_{\rm K}$, \textit{J}-\textit{K} colour magnitude diagram (figure \ref{mg_hyades}). Four objects, J0103+19, J0908+50, J1326$-$27 and J1343+29 are marked as being uncertain members to the moving group. They lie above the binary sequence in the colour magnitude diagram when the spectroscopic distance is used making them narrowly non members, however, when the moving cluster distance is used, J1326$-$27 and J1343+29 become more likely to be members and the remaining two objects as well as J1206+02 become less likely to be members.
It should be noted that J1553+15 is a binary object and also a T dwarf (making the spectroscopic distance incorrect). This object is marked with a diamond on figure \ref{mg_hyades}.\\
To illustrate the method we also show in figure \ref{mg_hyadesall} all the remaining proper motion dwarfs with their M$_{\rm K}$ calculated from their moving group distances. Since they are not members, they form a scattered distribution.

J1441-09 is believed to be a member of the Hyades moving group as stated by \citet{seifahrt05}. It was not selected as a member here as the ratio of moving group distance to spectral type distance was 1.47.  If the parallax distance is used, this ratio decreases to 1.35 which is still too high to be selected. $\Delta\theta$ however is only small indicating possible membership ($\approx$3$^{\circ}$). For the purposes of this study, it is not treated as a member.
For clarity figure \ref{mg_haydes} repeats figure \ref{mg_hyadesall} without the obvious non members. The \citet{bannister07} sequence is shown by a dashed line. Table \ref{hyades_table} contains more information about the suggested moving group members.

The following four objects are possible members and are marked by filled circles on figure \ref{mg_hyades}.\\
\textbf{J0103+19} is an L6 dwarf as identified by optical spectra \citep{kirkpatrick00}.\\
\textbf{J0908+50} is an L5 dwarf as identified by optical spectra by \citet{cruz03}. However \citet{knapp04} identify it as an L9$\pm$1 dwarf using infrared spectra.\\
\textbf{J1326-27} is an L5 dwarf (from optical spectrum) \citep{gizis02}. This is another object that was found not to be a member of the TW Hydrae association by \citet{gizis02}.\\
\textbf{J1343+39} is an L5 dwarf, identified by optical spectrum from \citet{kirkpatrick00}.\\

The remaining objects were selected as being probable members of the moving group and are plotted as encircled filled circles on figure \ref{mg_hyades}.\\
\textbf{J0228+25} is an L0 dwarf as identified by \citet{wilson03} using an infrared spectrum and by \citet{cruz03} using an optical spectrum. \\
\textbf{J1010-04} is an L6 which was discovered in 2003 by \citet{cruz03} and identified using an optical spectrum.\\
\textbf{J1045-01} is an L1 dwarf (from optical spectrum). \citet{gizis02} took a spectrum of this object to search for gravity features in the spectrum
 that would indicate youth and hence  
membership of the TW Hydrae association, however it was not found to be a member.\\
\textbf{J1047-18} is an L2.5 dwarf (from optical spectrum) \citep{martin99} and was originally discovered using the DENIS catalogue \citep{denis05}.\\
\textbf{J1207+02} is an L8 dwarf as measured from optical spectra by \citet{hawley02} and a T0 dwarf as measured from infrared spectra by \citet{burgasser06}. \\
\textbf{J1402+01} is an L1 dwarf discovered from the SDSS by \citet{hawley02}.\\
\textbf{J1404+46} is identified as an L0 dwarf from its optical spectrum \citep{cruz07}.\\
\textbf{J1407+12} is identified as an L5 dwarf by Reid et al (in prep).\\
\textbf{J2057+17} is identified as a L1 dwarf from its optical spectrum by \citet{kirkpatrick00}.\\
\textbf{J2107-03} is an L0 dwarf (from optical spectrum) \citep{cruz03}.\\
\textbf{J2130-08} is identified as an L1.5 dwarf by Kirkpatrick et al. (in prep).\\

\textbf{J1553+15} is identified as a T7 dwarf, and is also  a binary separated by 0.349'' \citep{burgasser06}. As the spectroscopic distance used here is 
only valid for dwarfs with spectral types from M6 to L8, the distance is incorrect and it is likely that this object is not actually a member. \citet{burgasser06b} 
calculate a distance for this object of 12$\pm$2 pc using the relationships of \citet{tinney03}. If this distance is used, a the ratio of moving cluster distance to spectral type distance is 1.31, which while close, is outside the limits of the selection criteria used here. The M$_{\rm K}$ of this object then becomes too faint to sit on the L$-$T transition sequence for this moving group. It is plotted as a filled diamond in figure \ref{mg_hyades}.\\

\begin{table*}
\caption{\label{hyades_table}Name $J$, $H$, $K$ magnitudes, $\Delta\theta$, d$_{mg}$/d$_{sp}$ and d$_{sp}$ for the potential Hyades moving group members discussed.}
\begin{center}
\begin{tabular}{l c c c c c c}
\hline
Name&$J$&$H$&$K$&$\Delta\theta$&d$_{mg}$/d$_{sp}$&d$_{sp}$\\
&&&&$^{\circ}$&pc\\
\hline
J0103+19  & 16.28 $\pm$ 0.07 & 14.89 $\pm$ 0.05 & 14.14 $\pm$ 0.05& 2.87 $\pm$ 2.68 & 1.09&28.62\\
J0228+25  & 13.83 $\pm$ 0.02 & 12.99 $\pm$ 0.02 & 12.47 $\pm$ 0.02& 3.78 $\pm$ 3.76 & 1.25&26.44\\
J0908+50  & 14.54 $\pm$ 0.02 & 13.47 $\pm$ 0.02 & 12.94 $\pm$ 0.02& 11.51 $\pm$ 2.06 & 0.80&15.91\\
J1010-04  & 15.50 $\pm$ 0.05 & 14.38 $\pm$ 0.03 & 13.61 $\pm$ 0.04& 7.54 $\pm$ 2.36 & 1.26&19.98\\
J1045-01  & 13.15 $\pm$ 0.02 & 12.35 $\pm$ 0.02 & 11.77 $\pm$ 0.02& 7.17 $\pm$ 1.39 & 1.04&17.09\\
J1047-18  & 14.19 $\pm$ 0.02 & 13.42 $\pm$ 0.02 & 12.89 $\pm$ 0.02& 6.61 $\pm$ 2.31& 1.17&22.10\\
J1207+02  & 15.57 $\pm$ 0.07 & 14.56 $\pm$ 0.06 & 13.98 $\pm$ 0.05& -8.77 $\pm$ 2.12 & 1.27&14.72\\
J1326-27  & 15.84 $\pm$ 0.07 & 14.74 $\pm$ 0.05 & 13.85 $\pm$ 0.05& 2.12 $\pm$ 2.17 & 0.89&28.93\\
J1343+39  & 16.16 $\pm$ 0.07 & 14.85 $\pm$ 0.05 & 14.14 $\pm$ 0.04& -1.38 $\pm$ 4.04 & 0.79&33.44\\
J1402+01  & 15.45 $\pm$ 0.06 & 14.65 $\pm$ 0.06 & 14.17 $\pm$ 0.07& 6.37 $\pm$ 2.72 & 0.78&49.08\\
J1404+46  & 14.33 $\pm$ 0.02 & 13.53 $\pm$ 0.02 & 13.05 $\pm$ 0.02& -9.35 $\pm$ 5.79 & 1.06&33.27\\
J1407+12  & 15.37 $\pm$ 0.05 & 14.34 $\pm$ 0.05 & 13.59 $\pm$ 0.04& -1.68 $\pm$ 3.33 & 1.20&23.31\\
J2057+17  & 15.96 $\pm$ 0.08 & 15.19 $\pm$ 0.08 & 14.49 $\pm$ 0.07& 5.82 $\pm$ 7.81 & 1.03&58.11\\
J2107-03  & 14.19 $\pm$ 0.02 & 13.44 $\pm$ 0.03 & 12.87 $\pm$ 0.02& -10.39 $\pm$ 4.4 & 1.17&31.22\\
J2130-08  & 14.13 $\pm$ 0.02 & 13.33 $\pm$ 0.03 & 12.81 $\pm$ 0.03& -5.94 $\pm$ 2.24 & 0.76&25.04\\
\hline
\end{tabular}
\end{center}
\end{table*}
This result indicates that many brown dwarfs may have escaped the Hyades cluster over time, as has been suggested, however they may have remained members of the Hyades moving group. If so, these objects may allow a study of the initial mass function of the Hyades to be made down to substellar masses.
A next step would be to measure parallaxes for these objects.
\subsection{Ursa Major}
The Ursa Major moving group has been estimated to have an age of between 300 Myr \citep{soderblom93} and 500$\pm$100 Myr \citep{king03}.\citet{castellani02} found an age of the group to be 400 Myr. The age of 400$\pm$100 Myr is adopted in this work.
The convergent point of the Ursa Major moving group is located at $\alpha$=20$^{\rm h}$18$^{\rm m}$.83, $\delta$=$-$34$^{\circ}$25'.8 (J2000, \citet{madsen02}).
Out of the 143 brown dwarfs examined, 4 appear to have the correct motion, distance and the correct direction of motion to be members of the Ursa Major moving group.

\begin{figure*}
\begin{center}
\scalebox{0.50}{\includegraphics[angle=270]{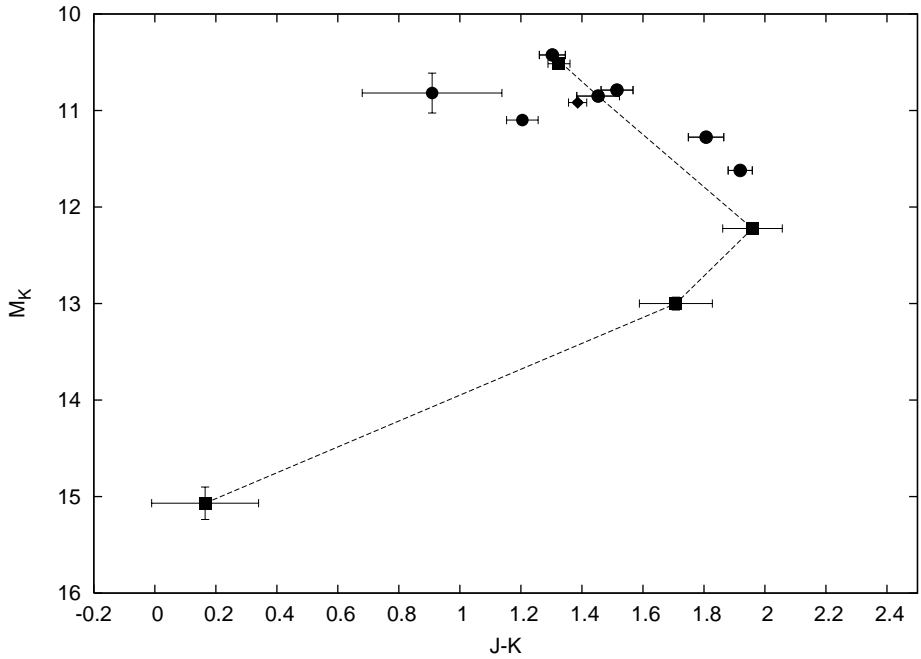}}
\caption{
\label{mg_ursa}
M$_{\rm K}$, $J$-$K$ colour magnitude diagram for the Ursa Major moving group. The members identified by \citet{bannister07} are plotted as filled squares. All of the objects that were selected as possible members are marked as filled circles and the selected members are marked by a ring around the filled circle. The diamond is the object J1017+13.The errors are poissonion and from the photometry only.The dashed line indicates the possible single star sequence.}
\end{center}
\end{figure*}

These objects are J0030+31, J1204+32, J1246+40 and J1550+14. Three objects had the correct motion, distance and direction however when placed on the colour magnitude diagram didn't fit the sequence. These objects are the binary J1017+13, J1147+02 and J1619+00, all of which lie too low on the main sequence to belong to the cluster. J1017+13
appears to be a possible member of the moving group however, it should be moved downwards to compensate for the fact that it is a binary, hence making it appear more likely to be a non member. It is plotted as a filled diamond on figure  \ref{mg_ursa}. The spectral types of these objects agree with their placement on the main sequence.\\
These objects are shown in figure \ref{mg_ursa} and their data in table \ref{ursatab}. \\
These objects were selected as being probable members of the moving group and are plotted as encircled filled circles on figure \ref{mg_ursa}.\\
\textbf{J0030+31} is an L2 (spectral type from optical spectrum) dwarf identified from 2MASS by \citet{kirkpatrick99}.\\
\textbf{J1204+32} is an L0 dwarf (spectral type from optical spectrum), with an estimated distance of 26.9 pc (from spectral type) \citep{cruz03}. \citet{wilson03} find a spectral type for this object of M9 from an infrared spectrum.\\
\textbf{J1239+55} is an L5 dwarf identified by its optical spectrum by \citet{kirkpatrick00}. It is also part of a resolved binary system.\\
\textbf{J1246+40} is an L4 dwarf (spectral type from optical spectrum) \citep{kirkpatrick00}. It has an estimated distance of 25 pc calculated from spectral type.\\
\textbf{1550+14} is an L2 dwarf (spectral type from optical spectrum) \citep{cruz07}.\\

\begin{table*}
\caption{\label{ursatab}Name $J$,$H$, $K$ magnitudes, $\Delta\theta$, d$_{mg}$/d$_{sp}$ and d$_{sp}$ for the potential Ursa Major moving group members discussed.}
\begin{center}
\begin{tabular}{l c c c c c c}
\hline
Name&$J$&$H$&$K$&$\Delta\theta$&d$_{mg}$/d$_{sp}$&d$_{sp}$\\
&&&&$^{\circ}$&pc\\
\hline
J0030+31  & 15.47 $\pm$ 0.05 & 14.61$\pm$0.05 & 14.02$\pm$0.047& 6.29$\pm$11.33 & 1.16&43.18\\
J1204+32  & 13.81$\pm$0.03 & 13.09$\pm$0.03 & 12.51$\pm$0.027& 8.02$\pm$12.30 & 1.13&26.18\\
J1239+55  & 14.71$\pm$0.02 & 13.56$\pm$0.03 & 12.79$\pm$0.027& 16.17$\pm$3.01 & 1.00&17.14\\
J1246+40  & 15.08$\pm$0.04 & 13.94$\pm$0.03 & 13.28$\pm$0.038& 12.26$\pm$5.41 & 0.72&25.17\\
J1550+14  & 14.77$\pm$0.04 & 13.79$\pm$0.03 & 13.26$\pm$0.034& 10.83$\pm$5.73& 0.72&31.23\\
\hline
\end{tabular}
\end{center}
\end{table*}
\subsection{Pleiades}
The Pleiades cluster is 125 Myr old and is situated at a distance of 130 pc \citep{stauffer98}.
As a cluster it has been studied in depth and has been found to contain many brown dwarfs \citep{casewell07, lodieu07a,bihain06,moraux03}. The Pleiades moving group has a convergent point of 85.04$^{\circ}$$\pm$3.67, $-$39.11$^{\circ}$$\pm$6.92 \citep{madsen02}. This convergent point is very close to that of many other moving groups such as  Alpha Persei(96$^{\circ}$.78$\pm$1.96, -23$^{\circ}$.27$\pm$3.67, \citet{madsen02}, 50 Myr \citet{lynga87}), Tucana/Horologium (30 Myr \citet{zuckerman04}) and the AB Dor moving group (50 Myr \citet{zuckerman04}) (see \citet{zuckerman04} for a review), and it has been theorised that many of these moving groups have a common origin \citep{ortega07}.

\begin{figure*}
\begin{center}
\scalebox{0.50}{\includegraphics[angle=270]{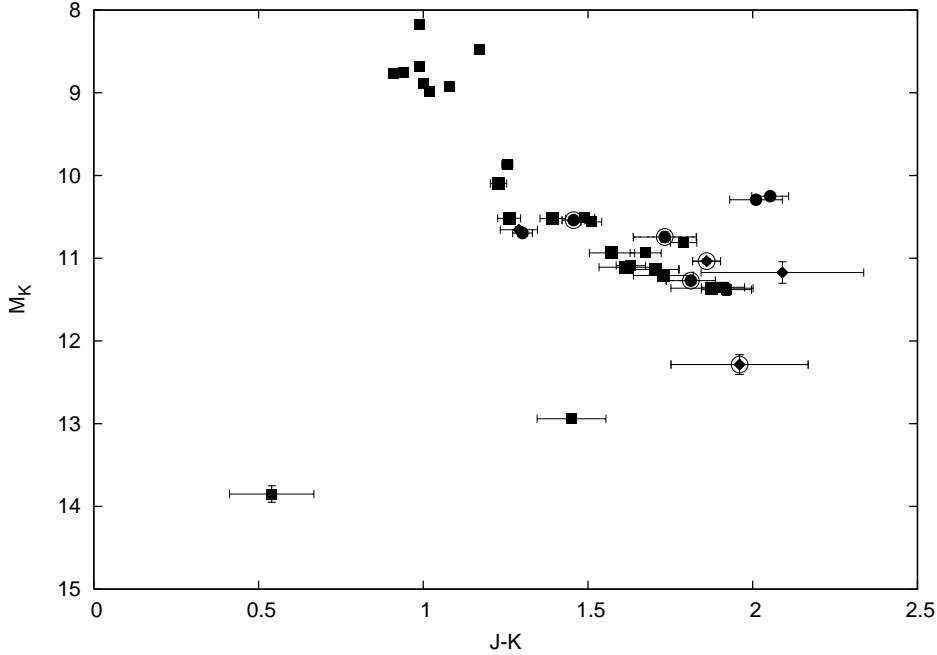}}
\caption{
\label{mg_plds}
M$_{\rm K}$, $J$-$K$ colour magnitude diagram for the Pleiades moving group. The cluster members members identified by \citet{casewell07}, \citet{lodieu07a}, \citet{moraux03} and \citet{bihain06} are plotted as filled squares.  All of the objects that were selected as possible members are marked as filled circles and the selected members are marked by a ring around the filled circle. The filled diamonds are objects that were selected as possible members that had measured parallax, and are outlined by a ring if considered a selected member.The errors are poissonion and from the photometry only.}
\end{center}
\end{figure*}

Three objects were selected to be potential members of the Pleiades moving group, and in addition to these three, one object with previously measured proper motion, and a parallax measurement also appears to fit the main sequence, while having the moving group distance and the parallax agree, as well as having motion in the correct direction to be a member.
These 4 objects were compared to the L-T dwarf sequence for the objects identified as being potential members of the Pleiades cluster by \citet{casewell07}. 
The spectral types of all of these objects are consistent with their places on the colour magnitude diagram as shown in figure \ref{mg_plds} and their data is in table \ref{pleiadestab}.\\
These objects were selected as members of the Pleiades moving group and are plotted as encircled filled points on figure \ref{mg_plds}.\\
\textbf{J0001+15} is an L4$\pm$1 dwarf as identified by \citet{knapp04} who used an optical spectrum.\\
\textbf{J1123+41} is an L2.5 dwarf discovered by \citet{kirkpatrick00}.\\
\textbf{J1552+29} has been identified as an L1 dwarf by \citet{wilson03}.\\
These two objects with known parallaxes were selected as members of the moving group and are plotted as filled, encircled diamonds on figure \ref{mg_plds}. \\
\textbf{GL417B} is a L4.5 dwarf discovered by \citet{kirkpatrick00}. It has a proper motion of 0.291$\pm$0.00072'' yr$^{-1}$ and a position angle of 238.67$\pm$0.14$^{\circ}$ \citep{perryman97}. This object has a parallax of 46.04$\pm$0.9 mas which corresponds to a distance of $\approx$21.72 pc \citep{perryman97}. This L dwarf was identified as a binary by \citet{kirkpatrick01} who had noted in \citet{kirkpatrick00} that it was in close proximity to GL417, a G type dwarf star. These two objects were found to have a common proper motion. It is noted in \citet{kirkpatrick01} that the age of this system is not inconsistent with that of the Pleiades (the age was calculated using the ratio of x$-$ray to bolometric luminosity vs B$-$V). 
Comparing Gl417A to the evolutionary models of \citet{girardi00}, using the measured parallax as above,of  shows that it lies to the left of the main sequence at this point, indicating that it is not a member of the moving group, however it is recorded as a variable star of the BY Dra type which may account for this difference. Further study would be needed of this object to prove a membership of the moving group.\\
\textbf{2MASSWJ2101154+175658} is an L7.5 dwarf also identified by \citet{kirkpatrick00}. \citet{vrba04} measure a proper motion of 0.2085$\pm$0.00037 `` yr$^{-1}$ and a position angle of 136.33$\pm$0.51$^{\circ}$ for it. It has a measured parallax of 30.14$\pm$3.42 mas which corresponds to a distance of $\approx$ 33.17 pc. \citet{gizis03} identified this L dwarf as a binary using the HST and suggest that both components must be brown dwarfs and the secondary must have a spectral type of later than L8. The separation is estimated to be 0.232''.\\

\begin{table*}
\caption{\label{pleiadestab}Name $J$, $H$, $K$ magnitudes, $\Delta\theta$, d$_{mg}$/d$_{sp}$ and d$_{sp}$ for the potential Pleiades moving group members discussed. The last two objects have parallax measurements and hence the last but one column is the ratio of moving group distance to parallax distance, not the spectral type distance.}
\begin{center}
\begin{tabular}{l c c c c c c}
\hline
Name&$J$&$H$&$K$&$\Delta\theta$&d$_{mg}$/d$_{sp}$&d$_{sp}$\\
&&&&$^{\circ}$&pc\\
\hline
J0001+15  & 15.52 $\pm$ 0.06 & 14.50 $\pm$ 0.05 & 13.71 $\pm$ 0.04& 9.45 $\pm$ 4.98 & 0.89&30.74\\
J1123+41  & 16.07 $\pm$ 0.07 & 15.08 $\pm$ 0.08 & 14.34 $\pm$ 0.05& 4.77 $\pm$ 7.62 & 0.88&52.41\\
J1552+29  & 13.47 $\pm$ 0.02 & 12.60 $\pm$ 0.02 & 12.02 $\pm$ 0.02& 12.20 $\pm$ 6.90 & 0.77&19.78\\
Gl 417B&14.58$\pm$0.03&13.50$\pm$0.03&12.72$\pm$0.03&5.00$\pm$0.14&0.94&21.72\\
2MASSWJ2101154+175658 &16.85$\pm$0.17&	15.86$\pm$	0.18&	14.89$\pm$0.12&9.48$\pm$0.51&0.67&33.18\\
\hline
\end{tabular}
\end{center}
\end{table*}

\section{Conclusions}
We have measured the proper motions for 143 dwarfs from the Dwarf Archive. From these measurements, we find 4 new common proper motion wide binary or multiple systems.
We also identify 8 high velocity dwarfs i.e. dwarfs with tangential velocities $\geq$ 100 kms$^{-1}$. These dwarfs also have bluer than average \textit{J}-\textit{K} colours.
We argue that these are probably thick disc objects with an age of order 10 Gyr. We estimate their luminosities which are $\approx$10$^{-4}$L/L$_{\odot}$. This suggests that they are probably very low luminosity stars rather than brown dwarfs. If so, they may be some of the dimmest stars found to date. 
Finally we have found 15 L dwarfs that are potential members of the Hyades moving group, 5 that are potential members of the Ursa Major moving group and 5 that are potential members of the Pleiades moving group. The next obvious step towards confirming membership of these groups is to measure parallaxes for these dwarfs. Parallaxes will allow accurate distances to be used to compare with the moving group distance. \textit{Spitzer} 3.5, 4.49, 5.73 and 7.87 micron magnitudes will also be valuable for a fuller understanding of the high velocity metal poor dwarfs.

\section*{Acknowledgements}
SLC, PDD, STH and NL were supported by PPARC for the duration of this work.
Observations were made  at the
 United Kingdom Infrared Telescope, which is operated by the Joint Astronomy Centre on behalf of the U.K. Particle Physics and Astronomy Research Council.
This publication makes use of data products from the Two Micron All Sky Survey, which is a joint project of the University of Massachusetts and the Infrared Processing and Analysis Center/California Institute of Technology, funded by the National Aeronautics and Space Administration and the National Science Foundation.
Research has benefited from the M, L, and T dwarf compendium housed at DwarfArchives.org and maintained by Chris Gelino, Davy Kirkpatrick, and Adam Burgasser.
This research has made use of NASA's Astrophysics Data System Bibliographic Services.
We thank K. Cruz for details of the proper motions from \citet{schmidt07}.

\label{lastpage}

\end{document}